\documentclass[aps,pre, twocolumn, groupedaddress]{revtex4-1}
\usepackage{graphicx}
\usepackage{dcolumn}
\usepackage{amsmath}    % need for subequations
\usepackage{amssymb}
\usepackage{bm} 
\usepackage{hyperref}
\usepackage{latexsym}
\usepackage{verbatim}
\usepackage{color}
\usepackage{setspace}
\usepackage{ulem}
\usepackage{siunitx}
\def\beq{\begin{equation}}
\def\eeq{\end{equation}}

\def\beq{\begin{equation}}                           
\def\eeq{\end{equation}}                           
\def\bea{\begin{eqnarray}}                           
\def\eea{\end{eqnarray}}        

\begin{document}
%%%%%%%%%%%%%%%%%%%%%%%%%%%%%%%%%%%%%%%%%%%%%%%%%%%
%                               TITLE & ABSTRACT
%%%%%%%%%%%%%%%%%%%%%%%%%%%%%%%%%%%%%%%%%%%%%%%%%%%
%Title of paper

\title{Ground state properties and exact thermodynamics of a 2-leg anisotropic spin ladder
system}
\author{Sk Saniur Rahaman$^{1}$}
\email{saniur.rahaman@bose.res.in}
%\affiliation{S. N. Bose National Centre for Basic Sciences, J D Block, Sector III, Salt Lake City, Kolkata 700106}
\author{Shaon Sahoo$^{2}$}
\email{shaon@iittp.ac.in}
\author{Manoranjan Kumar$^{1}$}
\email{manoranjan.kumar@bose.res.in}
%\affiliation{S. N. Bose National Centre for Basic Sciences, J D Block, Sector III, Salt Lake City, Kolkata 700106}
\affiliation{$^{1}$S. N. Bose National Centre for Basic Sciences, J D Block, Sector III, Salt Lake City, Kolkata 700106}
\affiliation{$^{2}$Department of Physics, Indian Institute of Technology, Tirupati, India}
\date{\today}
\begin{abstract}
We study a frustrated two-leg spin ladder with alternate isotropic Heisenberg and Ising rung exchange interactions, 
whereas, interactions along legs and diagonals are Ising-type. All the interactions in the ladder are anti-ferromagnetic 
in nature and induce frustration in the system. This model shows four interesting quantum phases: 
(i) stripe rung ferromagnetic (SRFM), (ii) stripe rung ferromagnetic with edge singlet (SRFM-E), 
(iii) anisotropic antiferromagnetic (AAFM), and (iv) stripe leg ferromagnetic (SLFM) phase. We construct a quantum 
phase diagram for this model and show that in stripe 
rung ferromagnet (SRFM), the same type of sublattice spins (either $S$ or $\sigma$-type
spins) are aligned in the same direction. Whereas, in anisotropic antiferromagnetic phase, both $S$ and $\sigma$-type of 
spins are anti-ferromagnetically aligned with each other, two nearest $S$ spins along the rung form an anisotropic
singlet bond whereas two nearest $\sigma$ spins form an Ising bond. In large Heisenberg rung exchange interaction limit, 
spins on each leg are ferromagnetically aligned, but spins on different legs are anti-ferromagnetically aligned. 
The thermodynamic quantities like $Cv(T)$, $\chi(T)$ and $S(T)$ are also calculated using the transfer matrix method 
for different phase. The magnetic gap in the SRFM and the SLFM can be notice from $\chi(T)$ and $Cv(T)$ curves. 
\end{abstract}
\maketitle
\setstretch{1.0}
%\begin{multicols}{2}
\section{Introduction}
\label{sec:introduction}
The study of quantum phase transitions in low dimensional spin systems has been a frontier 
area of research due to abundance of effective low-dimensional magnetic materials 
\cite{hutchings1979, park2007, mourigal2012, drechsler2007, dutton201224, dutton2012108, maeshima2003, 
sandvik1996, johnston1987, dagotto1996}
which exhibits a zoo of phases \cite{okamoto1992, haldan1982, srwhite1996, chitra1995, mkumar2015, soos2016, 
ckm1969, shastry1981, srwhite1994, chubukov1991, furkawa2012, zhitomirsky2010, parvej2017}. 
The confinement and interplay of exchange  interactions in  low dimensional 
systems like spin chains \cite{heilmann1978, hutchings1979, umegaki2015}, spin ladders 
\cite{sandvik1996, johnston1996, barnes1993, dagotto1996} or two dimensional systems \cite{manousakis1991, singh2010} 
can give rise to various interesting ground state (GS) properties 
\cite{mourigal2011, enderle2010, seidov2017, mkumar2013, mkumar2016, sirker2010, hamada1988}. 
Recently synthesized materials show that many of these spin-1/2 systems are frustrated even in one dimension (1D) 
\cite{mourigal2012, dutton201224, park2007, drechsler2007}, whereas the  
low dimensional systems can be either geometrically frustrated i.e. antiferromagnet Heisenberg 
spin-1/2  on a triangular lattice \cite{anderson1973, fazekas1974} or exchange interaction driven frustration such as 
1D spin-1/2 system interacting with nearest neighbor interaction $J_1$ and antiferromagnetic next nearest  neighbor  
exchange interaction $J_2$ \cite{mg1969, srwhite1996, chitra1995, okamoto1992, mkumar2010, tonegawa1987, sebastian1996}. 
Frustrated model Hamiltonians of one dimensional systems and 
zigzag geometry \cite{Korotin1999, Korotin2000} are extensively studied theoretically and GS of these 
systems have  exotic phases like spin liquid \cite{savary2017, dagotto1996}, 
dimer \cite{srwhite1996, chitra1995, mkumar2015, soos2016, ckm1969, shastry1981, srwhite1994, haldan1982}, 
spiral/non-collinear spin phase \cite{mkumar2015, soos2016, dmaiti2019}, ferromagnetic phase etc. 

Spin chains and ladders can also have anisotropic exchange interactions 
\cite{curely1986, strecka2003,rezania2015, cizmar2010, thielemann2009} 
and some spin chains can have alternate Heisenberg and Ising exchange interactions \cite{rojas2016}, 
whereas exchange along the leg is Ising type. The Heisenberg-Ising model has 
been explored by few groups \cite{verkholayak2012, verkholayak2013, rojas2016}.  
The simplest model on a ladder geometry studied by Rojas et. al. \cite{rojas2016} with alternate anisotropic Heisenberg 
($J_x$,$J_x$,$J_z$) and Ising type ($J_0$) rung exchanges and intraleg exchange interaction ($J_1$) 
gives interesting ground-state phase diagram with phases like frustrated phase 1 (FRU1), 
antiferromagnetic phase etc. in large and small ratio of Ising to Heisenberg exchange interactions ($J^z/J_1$) limits
respectively. This model also shows interesting sharp peak in specific heat. Verkholyak et. al. 
\cite{verkholayak2013} studied an anisotropic model with Heisenberg rung exchange interaction ($J_1$) and 
Ising-type leg exchange interaction ($J_2$) and diagonal exchange interaction ($J_3$). 
They showed that GS can exhibit different phases e.g. stripe leg (SL), stripe 
rung (SR), N\'eel and quantum paramagnetic (QPM) phases etc. in the phase diagram of $J_3$-$J_1$ plane and 
the field dependence behavior in this model are also studied \cite{verkholayak2012}.
There are other studies of Heisenberg branched chain model which show interesting GS behavior and plateau phase 
in the presence of external magnetic field \cite{karl2019}.  

The thermodynamical properties of the one or quasi-one dimensional quantum spin models with alternating 
isotropic and anisotropic units
are studied extensively in recent times \cite{rojas2016, sahoo2012}. In presence of alternate Heisenberg and 
Ising rung and Ising leg interaction, the two consecutive units of the Hamiltonian become commuting and in 
such cases, the exact thermodynamical properties of these systems can be calculated using transfer matrix method. For example, 
the susceptibility and other related quantities were calculated exactly for an anisotropic helical single-chain 
magnet $Fe_2Nb$ using transfer matrix method \cite{sahoo2012}.

In this paper, we study a general anisotropic Heisenberg-Ising model on ladder geometry with alternate 
Heisenberg and Ising exchange rung interactions, whereas the  exchange interactions along the leg and along the diagonal
of the ladder are Ising type as shown in Fig. \ref{model}. The system exhibits anisotropic antiferromagnetic (AAFM),
stripe rung ferromagnetic (SRFM),  stripe rung ferromagnetic-edge (SRFM-E) and stripe leg ferromagnetic (SLFM) phases.
We use exact diagonalisation method to calculate the GS properties upto 24 sites using Davidson algorithm 
\cite{davidson1975} for diagonalisation of the Hamiltonian matrix, whereas the 
thermodynamical properties are studied using the transfer matrix method \cite{huangstat}. 
The specific heat, magnetic susceptibility, entropy and average energy are studied in various phases.

This paper is divided into four sections, and in section \ref{sec:modelham} we discuss the model Hamiltonian. 
In section \ref{sec:results} results are discussed and is divided into four subsections. We summarise all the results and conclude 
in the section \ref{sec:summary}.  

\section{Model Hamiltonian}
\label{sec:modelham}
\begin{figure}[b]
\centering
\includegraphics[width=1.0\linewidth]{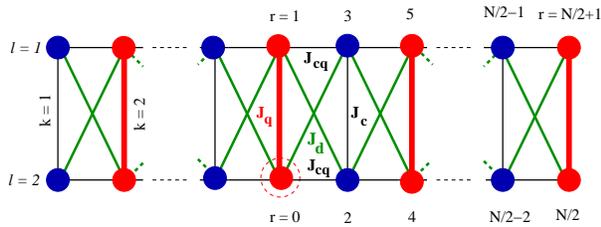}
\caption{(Color online). Schematic model diagram for spin configurations of a 2-leg anisotropic spin ladder system 
with $N (=4n)$ sites is shown. The interactions along the odd and even rungs are Ising and Heisenberg type respectively. 
Both along the legs and diagonals, the interactions are Ising-type. The indices $k$ ($1\le k \le 2n$) and $l$ (= 1 or 2) in the 
figure represent the rung and leg number; any spin in the anisotropic or isotropic rung is denoted by 
$\sigma_{k,l}$ or $S_{k,l}$ respectively. The variable $r$ is the distance from the reference spin shown within a dotted 
circle. }
\label{model}
\end{figure}  
We consider here a frustrated spin-1/2 ladder with alternating isotropic Heisenberg and Ising type interactions. 
For convenience, the system is divided into two sublattices A and B. The A sublattice has Heisenberg rung interaction $J_q$ 
while the B sublattice has Ising rung interaction $J_c$. The spins of two sublattices 
are connected by Ising-type interaction $J_{cq}$ along the legs and also by diagonal Ising-type interaction  $J_d$. 
Since for the B sublattice, only the $z$-component of spins appear in the Hamiltonian, we represent 
these spins by $\sigma$, whereas the other spins $S$ have all three components. 
The schematic diagram of the spin model is shown in Fig. \ref{model}.

\label{modelham}
The Hamiltonian for this system (having $4n$ sites) with open boundary 
condition (OBC) is given by $\mathbf{H} = \sum_{i=1}^{n-1} \mathbf{H_i} + \mathbf{H_e}$ where,
\begin{eqnarray} \label{eq:geometrical_general}
\centering
 &&  \mathbf{H_i}= J_{q} \vec{S}_{2i,1} \cdot \vec{S}_{2i,2}
    + \frac{J_c}{2} (\sigma_{2i-1,1}\sigma_{2i-1,2}+\sigma_{2i+1,1}\sigma_{2i+1,2}) \nonumber \\
 && + J_{cq} \left\{ S_{2i,1}^{z}(\sigma_{2i-1,1}+\sigma_{2i+1,1})+S_{2i,2}^{z}(\sigma_{2i-1,2}+\sigma_{2i+1,2}) 
           \right\} \nonumber \\
 && + J_{d} \left\{ S_{2i,1}^{z} (\sigma_{2i-1,2}+\sigma_{2i+1,2})+ S_{2i,2}^{z} (\sigma_{2i-1,1}+\sigma_{2i+1,1})
           \right\} \nonumber \\
 && + \frac{h}{2} \sum_{l=1}^2(2S_{2i,l}^{z}+\sigma_{2i-1,l} +\sigma_{2i+1,l}), ~~\text{and}  \\
&& \mathbf{H_e} = (\frac{J_c+h}{2}) \sum_{l=1}^2 \sigma_{1,l}\sigma_{1,l}+J_q\vec{S}_{2n,1} \cdot \vec{S}_{2n,2}  \nonumber \\
&&     + J_{cq}\sum_{l=1}^2 S_{2n,l}^z\sigma_{2n-1,l}+J_d(S_{2n,1}^z\sigma_{2n-1,2}+S_{2n,2}^z\sigma_{2n-1,1}) \nonumber \\
&& + \frac{h}{2}\sum_{l=1}^2 (2S_{2n,l}^z+\sigma_{2n-1,l}).
\end{eqnarray}
Here $\mathbf{H_e}$ is the part of the Hamiltonian representing the two edges. With the periodic boundary 
condition (PBC), $\mathbf{H_e}$ vanishes and the total Hamiltonian becomes $\mathbf{H} = \sum_{i=1}^{n} \mathbf{H_i}$ 
with appropriate reduction of values of site index, e.g. $\sigma_{2n+1,1} \equiv \sigma_{1,1}$. If our system is considered 
to be summation over $n$ geometrical units, then each unit is represented by the $\mathbf{H_i}$. It may be noted here that 
$[\mathbf{H_i},\mathbf{H_j}] = 0$ even for $j=i+1$.

For this work we consider $J_c = J_{cq} = 1$. The GS phase diagram of the system is studied here with respective
to the parameters $J_d$ and  $J_q$ (both positive).

\section{Results}
\label{sec:results}
\label{quantum_ground_state}
\begin{figure}[t]
\centering
\includegraphics[width=1.0\linewidth]{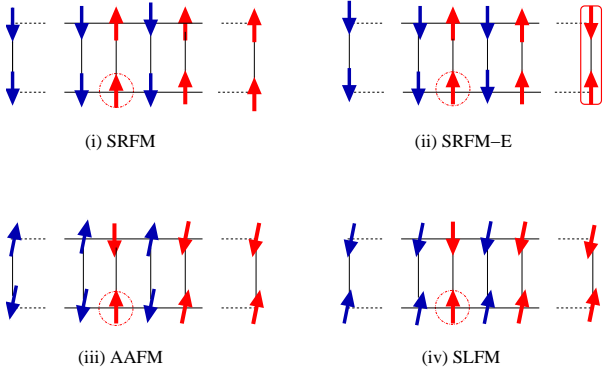}
\caption{(Color online). Spin arrangements in (i) stripe rung ferromagnetic (SRFM), (ii)  stripe rung ferromagnetic edge (SRFM-E), 
(iii) anisotropic antiferromagnetic (AAFM) and (iv) stripe leg ferromagnetic (SLFM) phases are shown. Arrows in the odd rungs 
(blue) and  even rungs (red)  represent $\sigma$-type and S type spins respectively. In subfigure (ii), the uncompensated dimer 
is shown in the box.} 
\label{allphases_diagram}
\end{figure}

In this section, fours phases are discussed in detail and to understand and characterize the phases and determine
their boundaries, we calculate various quantities like longitudinal $C^L(r)= <S^z_iS^z_{i+r}>$, 
transverse $C^t(r)=<(S_i^xS_{i+r}^x+S_i^yS_{i+r}^y)>$ correlations and 
energy crossovers. There are four major phases in the system: (i) stripe rung ferromagnet (SRFM) where the same 
type of sublattice 
spins (either S or $\sigma$-type spins) are aligned in the same direction, whereas other types are aligned 
along opposite direction as shown in Fig. \ref{allphases_diagram}.a. 
(ii) In stripe rung ferromagnetic-edge (SRFM-E) phase, bulk spins behave like SRFM phase, whereas the one of 
the edge spin pair ($S-S$) behaves like isolated singlet  as shown in Fig. \ref{allphases_diagram}.b  
and the GS is in $S^z_{tot}=1$ sector where $S^z_{tot}$ is the total $S^z$ for the entire ladder. 
(iii) In anisotropic antiferromagnetic (AAFM) phase, both  S and $\sigma$-type of spins are 
antiferromagnetically aligned with each other, two nearest $S$ spins along the rung form 
an anisotropic singlet bond, whereas two nearest $\sigma$ spins form an Ising bond 
as shown in Fig. \ref{allphases_diagram}.c.  The anisotropy of singlet bond decreases with 
increasing $J_q$ and spins are highly frustrated. (iv) In this phase, spins on each leg 
are ferromagnetically aligned but spins on other leg are antiferromagnetically aligned with each other 
(Fig. \ref{allphases_diagram}.d) and therefore this frustrated arrangement is called stripe leg ferromagnet (SLFM). 
\subsection{Quantum phase diagram}
\label{sec:phasediagram}
\label{phase_diagram}
\begin{figure}[t] 
\centering
\includegraphics[width=1.0\linewidth]{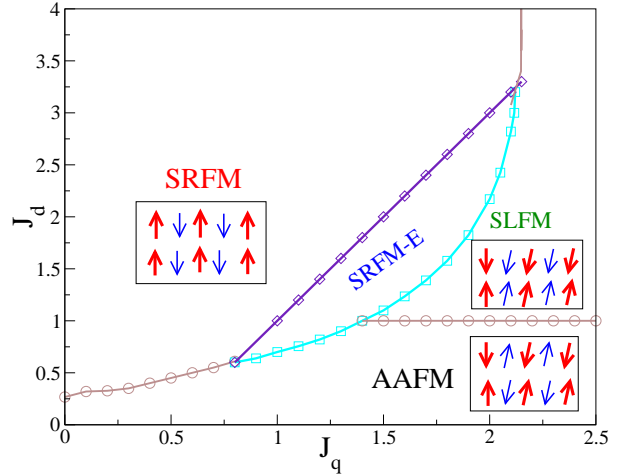}
\caption{(Color online). Quantum phase diagram of the Ladder with open boundary condition is shown. Spin arrangements of 
the AAFM, the SLFM, the SRFM phases are shown inside the boxes.}
\label{phase_diagram_n24}
\end{figure}
In Fig. \ref{phase_diagram_n24} the four phases, the SRFM, the SRFM-E, the AAFM and  the SLFM are shown separated by five 
phase boundaries for $N=24$, and we notice that the phase boundaries weakly depend on the system size. 
These phase boundaries are determined  based on energy crossovers and the 
correlation functions $C^L(r)$ and $C^T(r)$ by tuning $J_d$ and $J_q$. The large fraction of the phase space 
is covered by the SRFM phase and the AAFM phase has second largest contribution. It is interesting to note 
that the phase boundary of the AAFM and the SLFM is at $J_d/J=1$ for large $J_q$. Here, the bond order 
$<S_i.S_{i+1}>$  between the two $S$ spin along the rung form a perfect singlet dimer. The 
correlation  length in $C^L(r)$ shrinks to one unit cell, but this phase is restricted to only 
this phase boundary. The strong singlet dimers along the rung at $B$ type sublattice ($\sigma-\sigma$)
are formed on either sides of the phase boundary. 
\subsection{Ground state energy and excitation gap}
\label{sec:gs}
\begin{figure}[t]
\centering
\includegraphics[width=1.0\linewidth]{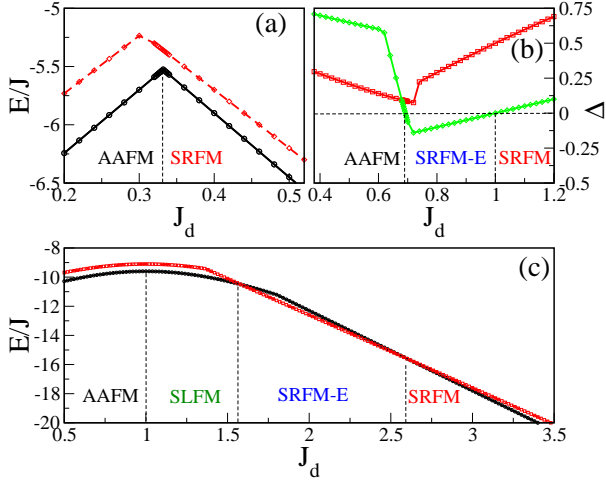}
\caption{(Color online). (a) Black solid line and red dashed line represent energies ($E/J$) in two respective spin sectors: 
$S^z_{tot}=0$ and $S^z_{tot}=1$ for $J_q=0.2$, (b) $\Delta$ is the energy gap for $J_q=1.0$. 
Red solid line represents the lowest energy gap in $S^z_{tot}=0$ sector and green solid line 
represents the energy gap of the lowest state in $S^z_{tot}=1$ sector from the lowest state in $S^z_{tot}=0$ sector. 
(c) Black solid line and red dashed line represent energies ($E/J$) in two respective spin sectors: 
$S^z_{tot}=0$ and $S^z_{tot}=1$ for $J_q=1.8$. ($E/J$) and $\Delta$ in all the subfigures are shown for system size N=24. }
\label{energy_crossover_n24}
\end{figure}

The GS energy $E_{GS}$ of the system is doubly degenerate in major part of the parameter space, 
and $E_{GS}$ and the lowest excited state in $S^z_{tot}=0$ and 1 sectors are analyzed  
as shown in Fig. \ref{energy_crossover_n24}. The lowest state energy in $S^z_{tot}=0$ and 1 sectors  are 
shown in  Fig. \ref{energy_crossover_n24}.a for $J_q=0.2$. The lowest energy $E_{GS}$ in 
$S^z_{tot}=0$ sector initially increases with $J_d$ due to enhancement in the frustration induced by $J_d$ and it 
starts to decrease again for $J_d >0.33$, as the $J_d$ becomes dominant and 
frustration decreases and system goes to the SRFM phase. The peak of $E_{GS}$ 
indicates the phase boundary. For small $J_q$ the phase transition from the AAFM to the SRFM seems 
to be sharp as derivative of $E_{GS}$ is discontinuous as shown in Fig. \ref{energy_crossover_n24} a. 
Whereas the change in  $E_{GS}$ is continuous for large $J_q$ as shown in Fig. \ref{energy_crossover_n24}, 
therefore phase transition seems to be second order.  In Fig. \ref{energy_crossover_n24}.b the lowest 
excited state in $S^z_{tot}=0 $ and the lowest state in $S^z_{tot}=1$ sector are shown with black and red color line-symbols 
for $J_q=1$ respectively. Negative value of red curve indicates the $S^z_{tot}=1$ as GS and the state appears 
because of a singlet dimer pair formation between edge $S-S$ spins, if the chain starts 
with $\sigma-\sigma$ spin pair (odd rungs) and ends with $S-S$ spin pair (even rungs) as considered in the system.  
In this case, a pair of  uncompensated ferromagnetically aligned $\sigma-\sigma$ pair gives rise to the $S^z_{tot}=1$ manifold. 
The boundaries for the SRFM-E is obtained by onset and end of the GS with $S^z_{tot}=1$ as shown in Fig. \ref{energy_crossover_n24}.b. 
In Fig. \ref{energy_crossover_n24}.c all four phases and their boundaries are shown for $J_q=1.8$. 
We notice that the maxima of doubly degenerate GS is the phase boundary between the AAFM and 
the SLFM phase, whereas, the onset and end of GS in the $S^z_{tot}=1$ is the phase boundary of the SRFM-E phase. 
In the SRFM phase, the GS is again in $S^z_{tot}=0$ sector. It is also evident from all three figures 
that $E_{GS}$ is continuous in large  $J_q$ limit. 

\subsection{Correlation functions}
\label{sec:correlation}
\begin{figure}[t]
\centering
\includegraphics[width=1.0\linewidth]{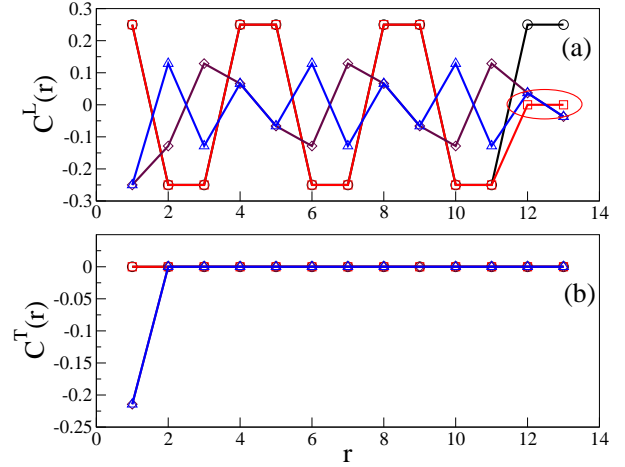}
\caption{(Color online). (a) Longitudinal and (b) Transverse Correlation plots are shown. Black, red, maroon and blue colors 
in both of the subfigures, represent four respective phases; SRFM ($J_q=0.2$,$J_d=2.0$), SRFM-E ($J_q=1.6$,$J_d=1.25$), 
AAFM ($J_q=2.0$,$J_d=0.4$) and SLFM ($J_q=2.0$,$J_d=1.6$).}
\label{allphase_corl}
\end{figure}
\begin{figure}[t]
\centering
\includegraphics[width=1.0\linewidth]{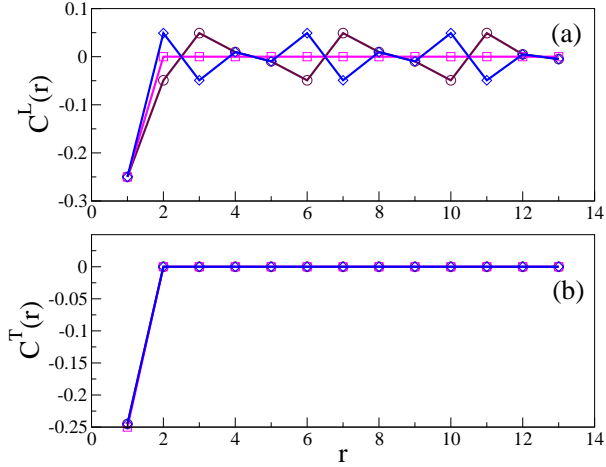}
\caption{(Color online). (a) Longitudinal and (b) Transverse Correlations for three phases are shown. Maroon,
magenta and blue colors in both of the subfigures are for AAFM ($J_q=2.0$, $J_d=0.8$), Perfect Dimer ($J_q=2.0$, $J_d=1.0$)
and SLFM ($J_q=2.0$, $J_d=1.2$) phases respectively.}
\label{dimer_corl}
\end{figure}

To understand the arrangement of spin in the GS, we study the two component: longitudinal $C^L(r)$ 
and transverse $C^T(r)$ correlations in four different phases as shown in Fig. \ref{allphase_corl}. 
The reference site is at the lower leg of sublattice A ($S$-type spin) at mid of the ladder and 
the arrangement of distance r is shown 
in Fig. \ref{model} . In the SRFM phase ({$J_q=0.2, J_d=2.0$}), the $C^L(r)$ shows long-range behavior and 
nearest neighbor along the rung is ferromagnetically aligned, whereas nearest neighbor 
along the leg is antiferromagnetically aligned. 
The $C^T(r)$ is zero for spins, therefore, GS is completely Ising like.   
In the SRFM-E ({$J_q=1.6, J_d=1.25$}), the correlation functions are same as that for the SRFM except at the boundary 
where the $C^L(r)$ goes to zero i.e the last pair of spins is decoupled from the ladder. The $C^T(r)$ 
is zero for all spins with respect to reference spin, but between edge rung spin pair $S-S$ it is -1/2. 
In the SLFM phase ({$J_q=2.0, J_d=1.6$}), the nearest rung spins are antiferromagnetically aligned, 
whereas along the leg nearest neighbor spins are ferromagnetically aligned. The nonzero value of 
$C^T(r)$ is restricted to nearest rung spin. However, in the limit {$J_q=2.0, J_d=0.4$} (AAFM phase), the $C^L(r)$ is 
long-range and both the nearest spins along the rung and along the leg are antiferromagnetically aligned. 
The $C^T(r)$ is restricted to the only nearest rung spin and the value $C^T(r=1)$ increases with $J_q$ as 
shown in Fig. \ref{allphase_corl}. It is also interesting to note that the long-range behavior in the  correlation 
$C^L(r)$ melts with increasing $J_q$. \\
The AAFM phase is interesting due to highly anisotropy correlations in the system and also the rapid variation 
in the correlation with $J_q$.  To our surprise, at $J_d=1$, two nearest spins along the rung ($S-S$ pairs) form
perfect singlet dimers, and the GS of the system behaves like product of  Ising and singlet dimers. To show the 
GS spin arrangement, $C^L(r)$ and $C^T(r)$ for $J_d= 0.8, 1$ and 1.2 for $J_q=2.0$ are plotted as a function 
of distance $r$ in Fig. \ref{dimer_corl}. We notice finite value of $C^L(r)$ and $C^T(r)$  
are restricted to nearest rung spin, whereas, $C^L(r)$ are non-collinear in nature in the neighborhood 
of $J_d=1$ for large $J_q$. For two values of $J_d=0.8$ and 1.2 for $J_q=2.0$, $C^L(r)$ shows non-collinear 
spin arrangement and the $C^T(r)$ is restricted to the same rung in A sublattice ($S$ spin) as shown 
in Fig. \ref{dimer_corl}.   
\subsection{ Exact thermodynamical properties}
\label{sec:thermodynamics}
The spin model Hamiltonian in Eq. \ref{eq:geometrical_general} have commuting bonds operators because Ising 
exchange interactions along the leg and diagonal of the ladder, therefore using transfer matrix method exact solution 
at finite temperature can be studied. In this paper, we study the low-temperature thermodynamical properties of our 
model using a suitably adapted transfer matrix method. Henceforth,  our transfer matrix calculations assume 
periodic boundary condition (PBC) and we will be using the full Hamiltonian without the edge part ($\mathbf{H_e}=0$). 
The Hamiltonian for a single geometrical unit (Eq.\ref{eq:geometrical_general}) can be reduced in the following manner:
\begin{eqnarray}\label{eq:master1}
\centering
      &&  \mathbf{H_i}=
           \frac {J_q}{2}\left(S_{2i,1}^{+}S_{2i,2}^{-}+S_{2i,2}^{+}S_{2i,1}^{-}\right)+
           J_q\left(S_{2i,1}^{z}S_{2i,2}^{z}\right) \nonumber \\ 
      &&   +a S_{2i,1}^{z}+b S_{2i,2}^{z}+c+d.
\end{eqnarray}
Here a, b, c, d can be written in terms of the parameters $J_{c}$, $J_{cq}$, $J_{d}$ and $h$, and the 
spin operator $\sigma$ (see in appendix \ref{appendix2}).
In the equation $S^{+}$, $S^{-}$ are the creation and annihilation operators respectively for spin S. 

Due to special construction of our model, we have $[\mathbf{H_i},\mathbf{H_j}]=0$ for any $i$ and $j$. 
This fact helps us to write 
the partition function of the total system as the trace of the $n$-th power of a small ($4\times4$) transfer matrix
(see the details in Appendix \ref{appendix1}). 
The partition function for $N (=4n)$ number of spins, $Q_N (\beta)=Tr(e^{-\beta \textbf{H}})$ with $\beta$ being the 
inverse temperature can be written as,
\begin{eqnarray}
\centering 
       && Q_{N}(\beta)= \lambda_{1}^{n}+\lambda_{2}^{n}+\lambda_{3}^{n}+\lambda_{4}^{n} ,   \nonumber  
\end{eqnarray} 
where four $\lambda$'s are the eigenvalues of the transfer matrix. 
If $\lambda_{1}$ is the largest eigenvalue then for large $N$, $ Q_{N}(\beta) = {\lambda_{1}}^{n}$ (see in appendix \ref{appendix4}). 
Using the partition function $Q_{4}(\beta)$ ($=\lambda_1$, partition 
function for a geometric unit), the thermodynamic quantities can be calculated using the following standard formulas: 
free energy (per geometrical unit) $F(T)=-k_{B}T  \log{Q_{4}  (\beta)}$, average energy 
$E(T)=k_{B}T^{2} \frac{d}{dT} \log{Q_{4} (\beta)}$, 
specific heat $C_{v}(T)=\left( \frac{\partial E(T)}{\partial T} \right)_{v}$, magnetization 
$M(T)=-\frac{\partial F(T)}{\partial h}$, magnetic susceptibility 
$\chi(T) =\frac{\partial M(T)}{\partial h}$, and entropy $S(T)=-\left( \frac{\partial F(T)} {\partial T} \right)$. 

In $T\rightarrow0$ limit, the largest eigenvalue $\lambda_1$ can be written as 
$\lambda_1 = e^{\frac{\beta (1+J_q(1+2\Delta_2))}{4}}+ e^{\frac{\beta (4J_d-J_q+3))}{4}}$, 
where $\Delta_2=\sqrt{1+4\frac{(1-J_d)^2}{J_q^2}}$ (see in appendix \ref{appendix4}). In the zero-temperature limit, the first exponential 
term in the expression of $\lambda_1$ dominates over the second exponential term in the regimes corresponding to the 
AAFM and the SLFM phases, while in the regime corresponding to the SRFM phase, the opposite happens.  
In this $T\rightarrow0$ limit, the free energy takes the following forms in the regimes corresponding 
to the SRFM and the AAFM phases respectively:
$F_{SRFM}=-\frac{4J_d-J_q+3}{4}$ and $F_{AAFM}=F_{SLFM}=-\frac{1+J_q(1+2\Delta_2)}{4}$. In this zero temperature 
limit, in all the three regimes, the entropy and the specific heat are found to be zero. These results match well 
with our numerical calculations using the full expression of $\lambda_1$ (see in appendix \ref{appendix3}).

\begin{figure}[t] 
\centering
\includegraphics[width=1.0\linewidth]{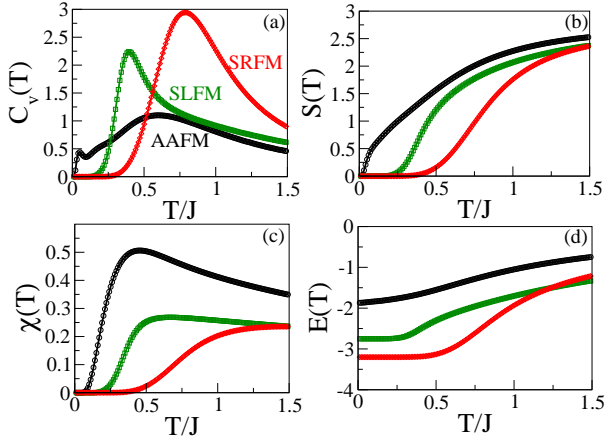}
\caption{(Color online). (a) Specific heat ($C_v(T)$), (b) Entropy ($S(T)$), (c) Magnetic susceptibility ($\chi (T)$) 
and (d) Average Energy ($E(T)$) plots are shown. Black, red and green curves in each of the subfigures, 
represent three phases: AAFM ($J_q=2.0$, $J_d=0.5$), SLFM ($J_q=2.0$, $J_d=2.5$) and SRFM ($J_q=0.2$, $J_d=2.5$) 
respectively.} 
\label{thermo}
\end{figure}

To understand the thermodynamic behavior at the non-zero temperatures, we calculate four thermodynamical 
quantities $E(T)$, $C_v(T)$, $S(T)$ and $\chi(T)$ for three parameter regimes and are shown in  Fig. \ref{thermo}. 
We use full expression of the largest 
eigenvalue $\lambda_1$ for this numerical calculation. It may be noted that the different ground state phases of 
the system, which were obtained with open boundary condition for the finite system sizes, may not have direct 
consequences in our low-temperature thermodynamic results as the thermodynamic calculations are done with periodic 
boundary condition for thermodynamically large system. Here, our main purpose of studying the thermodynamical 
quantities is to see how these quantities change across the parameter regimes of interest. 
The $C_v(T)$ of the three different phases show different features as shown in Fig. \ref{thermo}.a. In the 
AAFM region where $J_d$ is weak and $J_q$ is dominant, $C_v(T)$ shows a small peak near 
the $T\rightarrow 0$, which may be because of small gap due to small excitation gap in $S^{z}_{tot}=0$ sector, 
and then there is broad maxima at higher temperature, which is similar to the Heisenberg spin dimer system. 
The weak singlet dimer is formed along the rung of $S-S$ spins and that 
may give a broad peak at moderate temperature.  The $C_v(T)$  in  the SLFM phase shows very sharp 
peak and long tail, but have vanishing small value for $ T/J <0.09$ due to finite energy gap in 
the system. In the SRFM phase, this quantity is vanishingly small for $T/J <0.35$ due to large 
magnetic gap which makes the system to thermalise at higher temperature and a relatively 
higher peak at $T/J=0.75$. The entropy $S(T)$ is in some sense is
the measure of thermalisation, which in three different phases of the system are shown in Fig. \ref{thermo}.b.
In the AAFM phase, there is a small non-magnetic gap. Whereas, in other two phases $S(T)$ is vanishingly small                  
for $T/J <0.1$ due to large energy gap and thereafter it increases monotonically.

The magnetic susceptibility  $\chi(T)$ in these three phases are shown in Fig. \ref{thermo}.c and all the $\chi(T)$ have small 
values in all three phases for $T/J<0.1$. It has a broad maxima and small gap in the AAFM phase due to the formation of 
singlet dimer, and for breaking the weak singlet dimer it costs finite energy, therefore, singlet-triplet gap is finite. 
The $\chi(T)$ in the SRFM phase has dominant Ising interaction, therefore, there is a finite energy gap and 
sharp peak similar to the 1D Ising system. In the SLFM phase there is large magnetic gap as it requires breaking of 
strong rung interaction, and this leads to small $\chi(T)$ at low temperature and exponential increase in the $\chi(T)$. 
The average internal energy $E(T)$ shows a linear variation with
$T$ in the AAFM phase, but almost constant value of $E(T)$ for $T<0.09$ indicates the gap in the
SLFM phase as shown in Fig. \ref{thermo}.d. In the SRFM phase, variation of the energy is almost
constant for $T/J<0.35$ due to large energy gap and it varies linearly with $T$ thereafter.

\section{Summary and conclusion} 
\label{sec:summary}
In this paper, we consider a very general anisotropic Heisenberg-Ising model on ladder geometry 
with alternate Heisenberg and Ising exchange rung interactions, whereas the  exchange interactions 
along the leg and along the diagonal of the ladder are Ising type. We construct a quantum phase diagram 
of the model Hamiltonian in Sec.\ref{modelham}, and have shown that there are four quantum phases: (i) the AAFM, 
(ii) the SRFM, (iii) the SRFM-E and (iv) the SLFM which appear due to competing interactions and anisotropy 
in the system.
The GS is doubly degenerate and have finite magnetic gap in most of the parameter space and  to our surprise, 
exact dimer state along the rung in A sublattice (rungs with isotropic exchange interactions) appears for 
$J_d=1$ and large $J_q$ limit. However, weak dimer appears along rung of spin $S$ near to $J_d=1$.   

The thermal properties of this system are also studied analytically using the transfer matrix method. 
Four temperature dependent properties like specific heat $C_v(T)$, 
average internal energy $ E(T)$, entropy $S(T)$ and magnetic susceptibility $\chi(T)$ are studied 
in three different phases: the SRFM, the SLFM and the AAFM. In large $\frac{J_q}{J_d}$ regime (AAFM phase), 
$C_v$ shows a small peak at small $T$ due to a small excitation gap, whereas it has vanishingly 
small $\chi$ upto $T <0.09$ due to finite magnetic gap in the SLFM phase. Due to large excitation gap 
in the SRFM phase, all four quantities vanish for $T < 0.35$. 

In conclusion, we have studied a highly anisotropic model on a ladder geometry and the model Hamiltonian exhibits four 
interesting GS phases. The thermodynamic quantities like $Cv(T)$, $\chi(T)$, $E(T)$ and $S(T)$ are also studied 
using the transfer matrix method. This model may be realized in Cu or Vi based materials having magnetic interaction 
confined in ladder like geometry and the material should also have large anisotropy to ensure the Ising exchange. 

\section{Acknowledgements}
MK thanks DST India for a Ramanujan Fellowship SR/S2/RJN-69/2012. MK thanks SMST Department of IIT (BHU) for the 
hospitality during his visit. SS thanks SNBNCBS for supporting him under EVLP during his stay at the Centre when 
this work was started.

\newpage
\newpage
\section{Appendix}
\label{sec:appendix}
\label{appendix1}
         The partition function for $N$ sites, $Q_N(\beta)$ with Hamiltonian \textbf{$H$} can be written as- 
\begin{eqnarray}\label{eq:def_partition}
\centering
      &&  Q_{N}(\beta)= Tr \left( e^{-\beta \textbf{H}} \right)  %\nonumber
\end{eqnarray}
       where, Tr means trace of the matrix, $\beta= 1/\left(k_{B}T\right)$ and $k_{B}$ 
       is the Boltzmann constant. 
       Using explicit configuration basis for the system, Eq. \ref{eq:def_partition} is rewritten in the following form,
\begin{widetext}
\begin{eqnarray}%\label{eq:partition_basis0}
      &&Q_{N}(\beta)=\sum_{\{\sigma,S\}} 
     <\cdots, \sigma_{2i-1,1},\sigma_{2i-1,2},S_{2i,1},S_{2i,2},\cdots \mid 
         e^{-\beta \mathbf{H}} \mid        
     \cdots, \sigma_{2i-1,1},\sigma_{2i-1,2},S_{2i,1},S_{2i,2},\cdots >, \nonumber 
\end{eqnarray}
here the summation is over all possible configurations $\{\sigma,S\}$ of the system. For a given configuration, 
$|\cdots, \sigma_{2i-1,1},\sigma_{2i-1,2},S_{2i,1},S_{2i,2},\cdots >$ represents a basis state. Since for our system, the Hamiltonians corresponding
to different units commute with each other, we further get,

\begin{eqnarray}%\label{eq:partition_basis1}
      && Q_{N}(\beta)= \sum_{\sigma} <\cdots,\sigma_{2i-1,1},\sigma_{2i-1,2},\cdots \mid \prod_{i=1}^{n} 
		   \mathbf{T}_i \mid \cdots,\sigma_{2i-1,1},\sigma_{2i-1,2},\cdots> \nonumber
\end{eqnarray}

%\begin{eqnarray}%\label{eq:partition_basis1}
%      &&Q_N (\beta) =\sum_{\sigma} 
%     <\cdots,\sigma_{2i-1,1},\sigma_{2i-1,2},\cdots\mid
%         \prod_{i=1}^{n}\left( \sum_{\{S\}_i}<S_{2i,1},S_{2i,2}\mid e^{-\beta \mathbf{H_{i}}(\sigma,S)} 
%         \mid S_{2i,1},S_{2i,2}>\right)\mid \cdots,
%       \sigma_{2i-1,1},\sigma_{2i-1,2},\cdots > \nonumber,
%\end{eqnarray}
where $T_i=\sum_{\{S\}_i}<S_{2i,1},S_{2i,2}\mid e^{-\beta \mathbf{H_{i}}(\sigma,S)}\mid S_{2i,1},S_{2i,2}>$. Here the 
summation is over $\{S\}_i$ which represents all possible configurations of spins $S_{2i,1}$ and $S_{2i,2}$ (from the $i^{th}$ unit). 
It may be noted
that  $T_i$ does not contain the components of spin $S$ operators and it has only $\sigma$ 
variables, namely, $\sigma_{2i-1,1},\sigma_{2i-1,2}, \sigma_{2i+1,1}$ and $\sigma_{2i+1,2}$.
% Denoting this as $\mathbf{T}_i$, we get, 
%\begin{eqnarray}%\label{eq:partition_basis1}
%      && Q_{N}(\beta)= \sum_{\sigma} <\cdots,\sigma_{2i-1,1},\sigma_{2i-1,2},\cdots \mid \prod_{i=1}^{n} 
%           \mathbf{T}_i \mid \cdots,\sigma_{2i-1,1},\sigma_{2i-1,2},\cdots>. \nonumber
%\end{eqnarray}
\end{widetext}
This form is well-known with $\mathbf{T}_i$ being the transfer operator. Introducing identity operators 
$I = \sum_{\{\sigma\}_i}|\sigma_{2i-1,1},\sigma_{2i-1,2}> <\sigma_{2i-1,1},\sigma_{2i-1,2}|$ between successive $\mathbf{T}$ operators, we can finally write the
partition function as the trace of the $n$-th power of a small ($4\times4$) transfer matrix $\mathbf{P}$. We have, 
\begin{eqnarray}%\label{eq:partition_basis1}
      &&  Q_{N}(\beta)= Tr ( \mathbf{P}^n),\nonumber
\end{eqnarray}
       where $n$ is the number of geometrical units. The elements of the transfer matrix are given by
\begin{widetext}
\begin{eqnarray}\label{eq:pmat_elements}
      &&  P_{(\sigma_{2i-1,1},\sigma_{2i-1,2}),(\sigma_{2i+1,1},\sigma_{2i+1,2})} = 
<\sigma_{2i-1,1},\sigma_{2i-1,2}\mid \mathbf{T}_i \mid \sigma_{2i+1,1},\sigma_{2i+1,2}>
\end{eqnarray}
\end{widetext}

Before we construct and diagonalise the $\mathbf{P}$ matrix, we first need to carry out the trace over the configurations $\{S\}_i$ to find out the form of 
$\mathbf{T}_i$. Since $\mathbf{T}_i=\sum_{\{S\}_i}<S_{2i,1},S_{2i,2}\mid e^{-\beta \mathbf{H_{i}}(\sigma,S)}\mid S_{2i,1},S_{2i,2}>$, 
if we take the eigenstate 
basis of $\mathbf{H}_i$, we will get $\mathbf{T}_i$ as the summation over exponential of eigenvalues of $-\beta\mathbf{H}_i$. Next we calculate eigenvalues of 
$\mathbf{H}_i$ operator.  
 
\label{appendix2}
By considering, \\
      $  a=J_{d} \left( \sigma_{2i-1,2}^{z}+ \sigma_{2i+1,2}^{z} \right) +J_{cq}\left( \sigma_{2i-1,1}^{z}+
          \sigma_{2i+1,1}^{z}\right)+h $  \\
      $  b=J_{d} \left( \sigma_{2i-1,1}^{z}+ \sigma_{2i+1,1}^{z} \right) +J_{cq}\left( \sigma_{2i-1,1}^{z} +
          \sigma_{2i+1,1}^{z}\right)+h $  \\
      $  c=\frac{J_{c}}{2}\left(\sigma_{2i-1,1}^{z}\sigma_{2i-1,1}^{z} +
          \sigma_{2i+1,1}^{z}\sigma_{2i+1,2}^{z}\right) $ \\
      $  d=\frac{h}{2}\left(\sigma_{2i-1,1}^{z}+\sigma_{2i-1,2}^{z}+\sigma_{2i+1,1}^{z}+\sigma_{2i+1,2}^{z}\right) $, \\
         Hamiltonian (Eq. \ref{eq:master1} ) for the $i^{th}$ geometrical unit can be written as- 
\begin{widetext}
\begin{eqnarray} \label{eq:operators}
\centering
      &&  \mathbf{H_{i}}= \frac {J_q}{2}\left(S_{2i,1}^{+}S_{2i,2}^{-}+S_{2i,1}^{-}S_{2i,2}^{+}\right)+
           J_q\left(S_{2i,1}^{z}S_{2i,2}^{z}\right)   
        +a S_{2i,1}^{z}+b S_{2i,2}^{z}+c+d           \nonumber
\end{eqnarray} 
       By taking $ f=c+d $, we can write down the following Hamiltonian matrix in the eigenstate basis of  $S_{2i,1}^{z}S_{2i,2}^{z}$ operator, \\
\begin{equation}
{H_i}=\begin{pmatrix}
\frac{J_q}{4}+\frac{(a+b)}{2}+f	& 0		& 0		& 0		\\
0	& \frac{-J_q}{4}+\frac{(a-b)}{2}+f	& \frac{J_q}{2}	& 0		\\
0	& \frac{J_q}{2}	& \frac{-J_q}{4}-\frac{(a-b)}{2}+f	& 0		\\
0	& 0		& 0		& \frac{J_q}{4}-\frac{(a+b)}{2}+f	\\
\end{pmatrix}  .  \nonumber
\end{equation}
\end{widetext}
         The Hamiltonian matrix comes up with its four eigenvalues from three $S^z_{SS}$ sectors based on S-S pairs- \\
\textbf{(i) From $S^z_{SS}=1$ sector (formed by S-S pair)} \\
       $ \theta_1=(f+\frac{J_q}{4})+\frac{(a+b)}{2} $ \\
\textbf{(ii) From $S^z_{SS}=-1$ sector (formed by S-S pair)} \\
       $ \theta_2=(f+\frac{J_q}{4})-\frac{(a+b)}{2} $ \\
\textbf{(iii) From $S^z_{SS}=0$ sector (formed by S-S pair)} \\
       $ \theta_3=(f-\frac{J_q}{4}) + \frac{\sqrt{J_q^2+(a-b)^2}}{2} $ \\
       $ \theta_4=(f-\frac{J_q}{4}) - \frac{\sqrt{J_q^2+(a-b)^2}}{2} $. \\

We note that the eigenvalues ($\theta_k$) are functions of $\sigma$ variables, namely $\sigma_{2i-1,1},\sigma_{2i-1,2}, \sigma_{2i+1,1}$ and $\sigma_{2i+1,2}$. 
Using these eigenvalues, we rewrite $\mathbf{T}_i$ as,
\begin{eqnarray}
&&\mathbf{T}_i = \sum_{\{S\}_i}<S_{2i,1},S_{2i,2}\mid e^{-\beta \mathbf{H_{i}}(\sigma,S)}\mid S_{2i,1},S_{2i,2}> \nonumber \\
&&~~~= \sum_{k=1}^4 e^{-\beta \theta_k}.\nonumber
\end{eqnarray}

Without magnetic field (h=0), the Transfer Matrix ($\mathbf{P}$) takes the following form (using Eq. \ref{eq:pmat_elements})- \\

\begin{equation}
\mathbf{P}=\begin{pmatrix}
p	& q		& q		& r		\\
q	& s		& u		& q		\\
q	& u		& s		& q		\\
r	& q		& q		& p	\\
\end{pmatrix},     \nonumber  
\end{equation}   
here,
\begin{eqnarray}
   &&   p=2e^{\frac{-\beta}{4}}[Q^{-1} Cosh{\beta(1+J_d)}+Q Cosh(\frac{\beta J_q}{2})] \nonumber \\
   &&   q=2[Q^{-1} Cosh{\beta((1+J_d)/2)}+ Q Cosh(\frac{\beta J_q \Delta_{1}}{2})] \nonumber \\
   &&   r=2e^{\frac{-\beta}{4}}[Q^{-1}+Q Cosh(\frac{\beta J_q}{2})] \nonumber \\
   &&   s=2e^{\frac{\beta}{4}}[Q^{-1}+Q Cosh(\frac{\beta J_q \Delta_{2}}{2})] \nonumber \\
   &&   u=2e^{\frac{\beta}{4}}[Q^{-1}+Q Cosh(\frac{\beta J_q}{2})] \nonumber \\
   &&   \Delta_{1}=\sqrt{1+\frac{(1-J_d)^2}{J_q^2}}   \nonumber \\
   &&   \Delta_{2}=\sqrt{1+4\frac{(1-J_d)^2}{J_q^2}}   \nonumber \\
   &&    Q= e^{\frac{\beta J_q}{4}}.           \nonumber
\end{eqnarray}

\label{appendix3}
The above Transfer Matrix has $4$ simple eigenvalues $\lambda_{1}$, $\lambda_{2}$, $\lambda_{3}$, $\lambda_{4}$ as follows
\begin{eqnarray}
   && \mathbf{ \lambda_{1}}=\frac{(p+r+s+u)}{2} \nonumber \\
   &&  +\frac{\sqrt{(p+r+s+u)^{2}+16q^2-4(p+r)(s+u)}}{2} \nonumber \\
   && \mathbf{ \lambda_{2}}=\frac{(p+r+s+u)}{2} \nonumber \\
   &&  -\frac{\sqrt{(p+r+s+u)^{2}+16q^2-4(p+r)(s+u)}}{2} \nonumber \\
   && \mathbf{ \lambda_{3}}=(p-r) \nonumber \\
   && \mathbf{ \lambda_{4}}=(s-u). \nonumber 
\end{eqnarray}
It is to be noted that $\lambda_1$ is the largest eigenvalue here.

     In the special case with $T\rightarrow0$ limit, the largest eigenvalue can be approximated as- 
     $\lambda_{max}=(p+r+s+u)$. \\
\label{appendix4}
   Explicitly, we have, \\
\begin{eqnarray}
   &&   \mathbf{\lambda_{max}}=  \nonumber \\
   &&  (2e^{\frac{\beta (3J_q-1)}{4}}+e^{\frac{\beta (4J_d+3-J_q)}{4}}+ 
       5e^{\frac{\beta (1-J_q)}{4}}+e^{\frac{\beta (1+ J_q (1+2\Delta_2)}{4}}).  \nonumber 
\end{eqnarray}


\begin{thebibliography}{9}
%------------------------
      \bibitem{dutton2012108} S. E. Dutton, M. Kumar, M. Mourigal, Z. G. Soos, J. J. Wen, C. L. Broholm, N. H.
      Andersen, Q. Huang, M. Zbiri, R. Toft-Petersen, R. J. Cava, Phys. Rev. Lett.  $\mathbf{108}$, 187206 (2012).
      \bibitem{sandvik1996}A. W. Sandvik, E. Dagotto, and D. J. Scalapino, Phys. Rev. B $\mathbf{53}$, R2934 (1996).
     \bibitem{dagotto1996}E. Dagotto and T. M. Rice, Science $\mathbf{271}$, 618 (1996).
     \bibitem{maeshima2003} N. Maeshima, M. Hagiwara, Y. Narumi, K. Kindo, T. C. Kobayashi, and K. Okunishi, J. Phys.: 
     Cond. Matt. $\mathbf{15}$, 3607 (2003).
     \bibitem{johnston1987}D. C. Johnston, J. W. Johnson, D. P. Goshorn, and A. J. Jacobson, Phys. Rev. B $\mathbf{35}$, 219 (1987).
%-----------------------
\bibitem{hutchings1979} M. T. Hutchings, J. M. Milne, and H Ikeda, Journal of Physics C: Solid State Physics
            $\mathbf{12}$, L739 (1979).
% spin ladders
      \bibitem{park2007} C.L. Z. S. Park Y. J. Choi, and S. W. Cheon, Phys. Rev. Lett. $\mathbf{98}$, 057601 (2007).
       \bibitem{mourigal2012} Mourigal, M, Enderle, M, Fåk, B. and Kremer, R. K. and Law, J. M. and Schneidewind,
       A. and Hiess, A. and Prokofiev, A., Phys. Rev. Lett. $\mathbf{109}$, 027203 (2012).
     \bibitem{drechsler2007}S. L. Drechsler, O. Volkova, A. N. Vasiliev, N. Tristan, J. Richter, M. Schmitt, H. Rosner,
        J. Málek, R. Klingeler, A. A. Zvyagin, and B. Büchner, Phys. Rev. Lett. $\mathbf{98}$, 077202 (2007).
     \bibitem{dutton201224}S. E. Dutton, M Kumar, Z. G. Soos, C. L. Broholm, and R. J. Cava, J. Phys.: Cond. Matt.
        $\mathbf{24}$, 166001 (2012).
%------------------------
     \bibitem{mkumar2015}M. Kumar, A. Parvej, and Z. G. Soos, J. Phys.: Cond. Matt. $\mathbf{27}$, 316001 (2015).
      \bibitem{okamoto1992}K. Okamoto and K. Nomura, Phys. Lett. A  $\mathbf{169}$, 433 (1992).
      \bibitem{chitra1995}R. Chitra, S. Pati, H. R. Krishnamurthy, D. Sen, and S. Ramasesha, Phys. Rev. B $\mathbf{52}$, 6581 (1995). 
      \bibitem{soos2016}Z.G.Soos, A.Parvej and M.Kumar, J. Phys.: Condens. Matter $\mathbf{28}$, 175603(2016).
      \bibitem{ckm1969}C. K.  Majumdar  and  D. K.  Ghosh, J.  Math.  Phys. $\mathbf{10}$, 1388(1969).
      \bibitem{srwhite1994}S. R. White, R. M. Noack and D. J. Scalapino, Phys. Rev. Lett. $\mathbf{73}$, 886(1994).
      \bibitem{srwhite1996} S. R. White and I. Affleck, Phys. Rev. B $\mathbf{54}$, 9862 (1996).
      \bibitem{shastry1981}B. S. Shastry and B. Sutherland, Phys. Rev. Lett. $\mathbf{47}$, 964(1981).
      \bibitem{haldan1982}F. D. M. Haldane, Phys. Rev. B $\mathbf{25}$, 4925 (1982).
      \bibitem{chubukov1991}A. V. Chubukov, Phys. Rev. B $\mathbf{44}$, 4693 (1991).
      \bibitem{furkawa2012}S. Furukawa, M. Sato, S. Onoda, and A. Furusaki, Phys. Rev. B $\mathbf{86}$, 094417 (2012).
      \bibitem{zhitomirsky2010}M. E. Zhitomirsky and H. Tsunetsugu, Europhys. Lett. $\mathbf{92}$, 37001 (2010).
      \bibitem{parvej2017}A. Parvej and M. Kumar, Phys. Rev. B $\mathbf{96}$, 054413 (2017).
% low dimensional materials
      \bibitem{heilmann1978} I. U. Heilmann, G. Shirane, Y. Endoh, R. J. Birgeneau, and S. L. Holt, Phys. Rev. B $\mathbf{18}$,
          3530 (1978).
%      \bibitem{hutchings1979} M. T. Hutchings, J. M. Milne, and H Ikeda, Journal of Physics C: Solid State Physics
%            $\mathbf{12}$, L739 (1979).
      \bibitem{umegaki2015} I. Umegaki, H. Tanaka, N. Kurita, T. Ono, M. Laver, C. Niedermayer, C. Rüegg, S. Ohira-Kawamura,
             K. Nakajima, and K. Kakurai, Phys. Rev. B $\mathbf{92}$, 174412 (2015).
%2 dimensional ladder systems
     \bibitem{johnston1996}D. C. Johnston, Phys. Rev. B $\mathbf{54}$, 13009 (1996).
     \bibitem{barnes1993}T. Barnes, E. Dagotto, J. Riera, and E. S. Swanson, Phys. Rev. B $\mathbf{47}$, 3196 (1993).
% 2 dimensional systems
     \bibitem{manousakis1991}E. Manousakis, Rev. Mod. Phys. $\mathbf{63}$, 1 (1991).
     \bibitem{singh2010}Y. Singh and P. Gegenwart, Phys. Rev. B $\mathbf{82}$, 064412 (2010).
% gs phases
     \bibitem{mourigal2011} M. Mourigal, M. Enderle, R. K. Kremer, J. M. Law, and B. Fåk, Phys. Rev. B $\mathbf{83}$, 100409 (2011).
      \bibitem{enderle2010} M. Enderle, B. Fåk, H. J. Mikeska, R. K. Kremer, A. Prokofiev, and W. Assmus,
         Phys. Rev. Lett. $\mathbf{104}$, 237207 (2010).
      \bibitem{seidov2017} Z. Seidov, T. P. Gavrilova, R. M. Eremina, L. E. Svistov, A. A. Bush, A. Loidl
         and H. A. Krug von Nidda, Phys. Rev. B $\mathbf{95}$, 224411 (2017).
%zoo of phases
      \bibitem{mkumar2013}M Kumar, S. E. Dutton, R. J. Cava, and Z. G. Soos, J. Phys.: Cond. Matt. $\mathbf{25}$, 136004 (2013).
      \bibitem{mkumar2016}M. Kumar, A.Parvej, and Z. G Soos, J. Phys.: Cond. Matt. $\mathbf{28}$, 175603 (2016).
      \bibitem{sirker2010}J. Sirker, Phys. Rev. B $\mathbf{81}$, 014419 (2010).
      \bibitem{hamada1988}T. Hamada, J. Kane, S. Nakagawa, and Y. Natusume,J. Phys. Soc. Jpn. $\mathbf{57}$, 1891 (1988).
%triangular lattic
      \bibitem{anderson1973}P. Anderson, Materials Research Bulletin $\mathbf{8}$, 153 (1973).
      \bibitem{fazekas1974}P. Fazekas and P. W. Anderson, The Philosophical Magazine: A Journal of Theoretical Experimental and
                Applied Physics $\mathbf{30}$, 423 (1974).
% J1-J2 spin chain
      \bibitem{mg1969}C. K. Majumdar and D. K. Ghosh, J. Math. Phys. $\mathbf{10}$, 1399 (1969).
      \bibitem{sebastian1996}Sebastian Eggert. Phys. Rev. B $\mathbf{54}$, R9612 (1996).
      \bibitem{mkumar2010}M. Kumar, S. Ramasesha, and Z. G. Soos, Phys. Rev. B $\mathbf{81}$,054413 (2010).
      \bibitem{tonegawa1987} T. Tonegawa and I. Harada, J. Phys. Soc. Jpn. $\mathbf{56}$, 2153 (1987).
%2 dimensional zigzag ladder
     \bibitem{Korotin1999} M. A. Korotin, I. S. Elfimov, V. I. Anisimov, M. Troyer, and D. I. Khomskii,
       Phys. Rev. Lett. $\mathbf{83}$, 1387 (1999).
     \bibitem{Korotin2000} M. A. Korotin, V. I. Anisimov, T Saha-Dasgupta, and I Dasgupta, Journal of
	Physics: Cond. Matt. $\mathbf{12}$, 113 (2000).
% spin liquid 
      \bibitem{savary2017} Lucile Savary and Leon Balents, Rep. Prog. Phys.  $\mathbf{80}$, 016502 (2017).
% dimer
%      \bibitem{srwhite1996}S. R. White and I. Affleck, Phys. Rev. B $\mathbf{54}$, 9862(1996).
%      \bibitem{chitra1995}R.  Chitra,  S.  Pati,  H.  R.  Krishnamurthy,  D.  Sen,  and  S.Ramasesha, Phys. Rev. B $\mathbf{52}$, 6581(1995). 
%      \bibitem{mkumar2015}M. Kumar, A. Parvej and Z. G. Soos, J. Phys.: Condens. Matter $\mathbf{27}$, 316001(2015).
      \bibitem{dmaiti2019} D. Maiti, M. Kumar, Phys. Rev. B  $\mathbf{100}$, 24511 (2019).

%anisotropic magnets studied
%      \bibitem{dupas1982} Dupas. C,  Renard. J.P.,  Seiden. J,  Cheikh-Rouhou, Phys. Rev. B $\mathbf{25}$ 3261 (1982).
%      \bibitem{seiden1983} Seiden. J, J. Physique Lett. $\mathbf{44}$ 947 (1983).
%      \bibitem{verdaguer1984} Verdaguer. M., Gleizes. A., Renard. J.P., Seiden. J, Phys. Rev. B $\mathbf{29}$, 5144 (1984).
% anisotropy exchange
      \bibitem{curely1986} Cur\'ely. J, Georges. R, Drillon. M, Phys. Rev. B $\mathbf{33}$, 6243 (1986).
      \bibitem{strecka2003}Stre\ifmmode \check{c}\else \v{c}\fi{}ka. J, Michal Jaˇsˇcur. M,  J. Phys.: Cond. Matt. 
       $\mathbf{15}$, 4519 (2003).
      \bibitem{rezania2015}H. Rezania, Journal of Magnetism and Magnetic Materials $\mathbf{388}$, 68 (2015).
      \bibitem{cizmar2010}E. \ifmmode \check{C}\else \v{C}\fi{}i\ifmmode \check{z}\else \v{z}\fi{}m\'ar, E. and Ozerov, M 
       and Wosnitza, J. and Thielemann, B. and Kr\"amer, K. W. and R\"uegg, Ch. and Piovesana, O. and 
       Klanj\ifmmode \check{s}\else \v{s}\fi{}ek, M. and Horvati\ifmmode \acute{c}\else \'{c}\fi{}, M. and Berthier, C 
       and Zvyagin, S. A, Phys. Rev. B $\mathbf{82}$, 054431 (2010).
      \bibitem{thielemann2009} Thielemann, B. and R\"uegg, Ch. and Ronnow, H. M. and L\"auchli, A. M. and Caux, 
        J. S. and Normand, B. and Biner, D. and Kr\"amer, K. W. and G\"udel, H.-U. and Stahn, J. and Habicht, 
        K. and Kiefer, K. and Boehm, M. and McMorrow, D. F. and Mesot, J, Phys. Rev. Lett. $\mathbf{102}$, 107204 (2009).
% anisotropic spin ladder system
%     \bibitem{wesel2017}Stefan Wessel, Bruce Normand, Frédéric Mila and Andreas Honecker, SciPost Phys. $\mathbf{3}$, 005 (2017).
      \bibitem{rojas2016}Onofre Rojas and J. Stre\ifmmode \check{c}\else \v{c}\fi{}ka and S.M. {de Souza}, 
       Solid State Communications $\mathbf{246}$, 68 (2016).
      \bibitem{verkholayak2013}T. Verkholyak, J. Strečka, Cond. Matt. Phys. $\mathbf{16}$, 13601 (2013).
      \bibitem{verkholayak2012} T. Verkholyak and J. Streka, J. Phys. A: Math. Theor. $\mathbf{45}$,  305001 (2012).
% magnetic plateus
       \bibitem{karl2019}Karl'ov\'a, Katar\'{\i}na and Stre\ifmmode \check{c}\else \v{c}\fi{}ka, Jozef and Lyra, Marcelo L,
        Phys. Rev. E. $\mathbf{100}$, 042127 (2019).
% Transfer Matrix Method
      \bibitem{sahoo2012}S. Sahoo, J. P. Sutter, and S. Ramasesha, Journal of Statistical Physics $\mathbf{147}$, 181193 (2012).
%     \bibitem{venkataktrishnan}Venkatakrishnan, T.S, Sahoo. S, Bréfuel. N, Duhayon. C, Paulsen. C, Barra, A. L., Ramasesha, 
%      S. Sutter, J.P, J. Am. Chem. Soc. $\mathbf{132}$, 6047 (2010).      
%%%%%%%%%%%%%%%%%%%%%%%%%%%%%%%%%%%%%%%%%%%%%%%%%%%%%%
% Exact Diagonalisation
      \bibitem{davidson1975}Davidson, Ernest R., Journal of Computational Physics, $\mathbf{17}$, 87 (1975).
% Stat Mech Book
      \bibitem{huangstat}Kerson Huang, ISBN-13 : 978-8126518494
%2 dimensional ladder systems
%      \bibitem{Dagotto1999}E. Dagotto, Rep. Prog. Phys. $\mathbf{62}$, 1525 (1999).
%      \bibitem{Mikeska}H.J. Mikeska and A. K. Kolezhuk, ISBN 978-3-540-40066-0.
%      \bibitem{hiroi1991}Z. Hiroi, M. Azuma, M. Takano, and Y. Bando, J. Solid State Chem. $\mathbf{95}$, 230 (1991).
%      \bibitem{kim2000} E. H. Kim, G. Fáth, J. Sólyom, and D. J. Scalapino, Phys. Rev. B $\mathbf{62}$, 14965 (2000).
%      \bibitem{ameida2007}J. Almeida, M. A. Martin-Delgado, and G. Sierra, Phys. Rev. B $\mathbf{76}$, 184428 (2007).
%      \bibitem{volkova2012}O. S. Volkova, I. S. Maslova, R. Klingeler, M. Abdel-Hafiez, Y. C. Arango, A. U. B. Wolter, V.
%	Kataev, B. Büchner, and A. N. Vasiliev, Phys. Rev. B $\mathbf{85}$, 104420 (2012).
%      \bibitem{badrtdinov2016} D. I. Badrtdinov, O. S. Volkova, A. A. Tsirlin, I. V. Solovyev, A. N. Vasiliev, and V. V.
%	Mazurenko, Phys. Rev. B $\mathbf{94}$, 054435 (2016).
% anisotropic chain-magnets
%      \bibitem{bogani2008} Bogani, L., Vindigni, A., Sessoli, R., Gatteschi, D.: J. Mater. Chem. $\mathbf{18}$, 4750 (2008).
%      \bibitem{miyasaka2009} Miyasaka, H.,  Julve, M., Yamashita, M.,  C\'lerac,  R.: Inorg. Chem. $\mathbf{48}$, 3420 (2009).
%      \bibitem{coulon2004} Coulon, C., C\'lerac, R., Lecren, L., Wernsdorfer, W., Miyasaka, H.,
%       Phys. Rev. B $\mathbf{69}$, 132408 (2004).
%      \bibitem{vindigni2008} Vindigni, A., Inorg. Chim. Acta $\mathbf{361}$, 3731 (2008).
%      \bibitem{Sun2010} Sun, H. L., Wang, Z. M., Gao, S., Coord. Chem. Rev. $\mathbf{254}$, 1081 (2010).

%
\end{thebibliography}
\end{document}